\documentclass[12pt]{iopart}
\usepackage{iopams} 
\usepackage[normalem]{ulem}
\usepackage{epsfig}
\usepackage{graphics}
\usepackage{lipsum}
\usepackage{sidecap}
\usepackage{mathtools}
\usepackage{graphicx} 

\begin{document}
\title{Large Amplitude Spin-Hall Oscillations due to Field-like Torque}
\author{R. Arun$^{1}$,R. Gopal$^{2}$,V.~K.~Chandrasekar$^{2}$, and M.~Lakshmanan$^1$}
\address
{
$^{1}$Department of Nonlinear Dynamics, School of Physics, Bharathidasan University, Tiruchirapalli-620024, India\\
$^{2}$Centre for Nonlinear Science \& Engineering, School of Electrical \& Electronics Engineering, SASTRA Deemed University, Thanjavur- 613 401, India. \\
}
\ead{arunbdu@gmail.com,gopalphysics@gmail.com,chandru25nld@gmail.com,lakshman.cnld@gmail.com}

\date{\today}
	
\begin{abstract}
Large amplitude spin-Hall oscillations are identified in a ferromagnetic layer with two perpendicular in-plane easy axis in the presence of field-like torque without any polarizer and external field.  The analytical study confirms the possibility of oscillations in the presence of field-like torque. The investigation shows that the oscillation frequency can be tuned from $\sim$2 GHz to $\sim$80 GHz by current and enhanced by field-like torque. Further, the enhancement of frequency along with the Q-factor by current and field-like torque is also observed.
\end{abstract}

\vspace{2pc}

\submitto{Journal of Physics: Condensed Matter}

\maketitle

\section{Introduction}
The spin-transfer torque\cite{slon:1996,berger:1996} exerted by the spin-polarized electrons which arise due to pinned polarizers is the well-known source for the magnetization switching and magnetization oscillations\cite{katine,kiselev,grollier,rippard,tani,subash:2013,subash:2015,gopal:2019}. These oscillations arise due to the application of electric current in nanoscaled ferromagnetic/nonmagnetic layers. Another source of spin-polarized electrons which has been identified in such nanoscaled ferromagnetic materials is the spin-Hall effect (SHE)\cite{dyak:1971,hirs:1999}. It generates spin currents out of electric current that passes through the nonmagnetic heavy metal layer and these spin currents can be effectively used to manipulate the magnetization of a ferromagnetic layer attached to the heavy metal layer\cite{dyak:1971,hirs:1999,kato:2004,chen}. 

The spin current formation is achieved  when a current passes along the $x$-direction through a nonmagnetic heavy metal layer having high spin-orbit coupling, where the electrons which are polarized along the positive and negative $y$-directions are scattered along the positive and negative $z$-directions respectively. In a spin-Hall system where a ferromagnetic layer is placed on the nonmagnetic layer, the spin-transfer torque exerted by the incoming spin-polarized electrons, namely spin-Hall torque, exhibits magnetization dynamics in the ferromagnetic layer\cite{ando,miron,miron1}.  The direction of the SHE is determined geometrically and its magnitude indicates a different angular dependence in the ferromagnetic multilayer than the spin-torque\cite{ando}. Experimental  studies on magnetization switching and steady state precession are reported and the threshold current formula was derived for the switching of a perpendicular magnetization by the SHE in Refs.\cite{yang,liu,lee}. Further, spin-Hall oscillations and synchronization in spin-Hall oscillators have also been widely studied both theoretically and experimentally\cite{demi:2014,sira:2015,puli:2016,liu:2013,tani:2017,tani:2018,liu:2012,knap}.   Particularly, Liu $et~al$. have observed the SHE driven magnetization oscillations with frequency around 1.6 GHz in magnetic tunnel junction that consists of in-plane magnetized free layer of 1.5 nm thickness in the presence of external magnetic field\cite{liu:2012}.

It has been shown that the magnetization oscillations are possible in the spin-Hall system when placing an additional tilted polarizer\cite{zhou:2008,zhou:2009,zhou:2009a,arun:2020,bhoom:2018} above or below the ferromagnetic layer along with a spacer\cite{tani:2017,tani:2018}. The spin-Hall effect has been  utilized for self-oscillations in an in-plane magnetized system under the effect of applied magnetic field\cite{liu:2012}.  On the other hand, field-like torque\cite{zhang:2002,spiro:2003} has been identified as a source for self-oscillations without an external field in spin-valve systems\cite{tani:2014,guo:2015}. Also, it has been studied that the external magnetic field is essential for the out-of-plane magnetization oscillations in a spin-Hall system\cite{tani:2015}. Moreover, the need for the applied magnetic field has been severely limited for many possible applications such as the microwave oscillations,  magnetization switching by spin-orbit torque, and  memory and logic\cite{liu:2019}.

In this paper, we investigate the oscillations of magnetization in the spin-Hall system associated with field-like 
torque in the absence of any polarizer and external magnetic field.  Recently, Eden $et~al.$ have experimentally confirmed that two  in-plane easy axis anisotropies can be formed in a single layer at perpendicular directions by the oblique deposition method\cite{eden}. Also, Zhou $et~al.$ have deposited YIG ferrimagnetic film with two in-plane perpendicular easy axes and identified  magnetization switching between them\cite{Zhou:2020}.  We show the emergence of oscillations by changing the magnitude of two perpendicular in-plane easy axis anisotropies\cite{eden,Zhou:2020}  by solving the associated Landau-Lifshitz-Gilbert (LLG) equation  with the spin-Hall torque numerically and the expression for frequency is derived  analytically.  We also observe an impressive enhancement of the microwave oscillation frequency in the range of 2 GHz to 80 GHz, depending on the magnitudes of the in-plane easy axis anisotropies.

The paper is organized as follows: Sec.II addresses the model for the spin-Hall system and the governing LLG equation. In Sec.III, the existence of oscillations and the variation of oscillation frequency and power by current, field-like torque { and anisotropy}  are discussed.  Finally, Sec.IV presents the concluding remarks.

\section{Model}
\begin{figure}[htp]
	\centering\includegraphics[angle=0,width=1\linewidth]{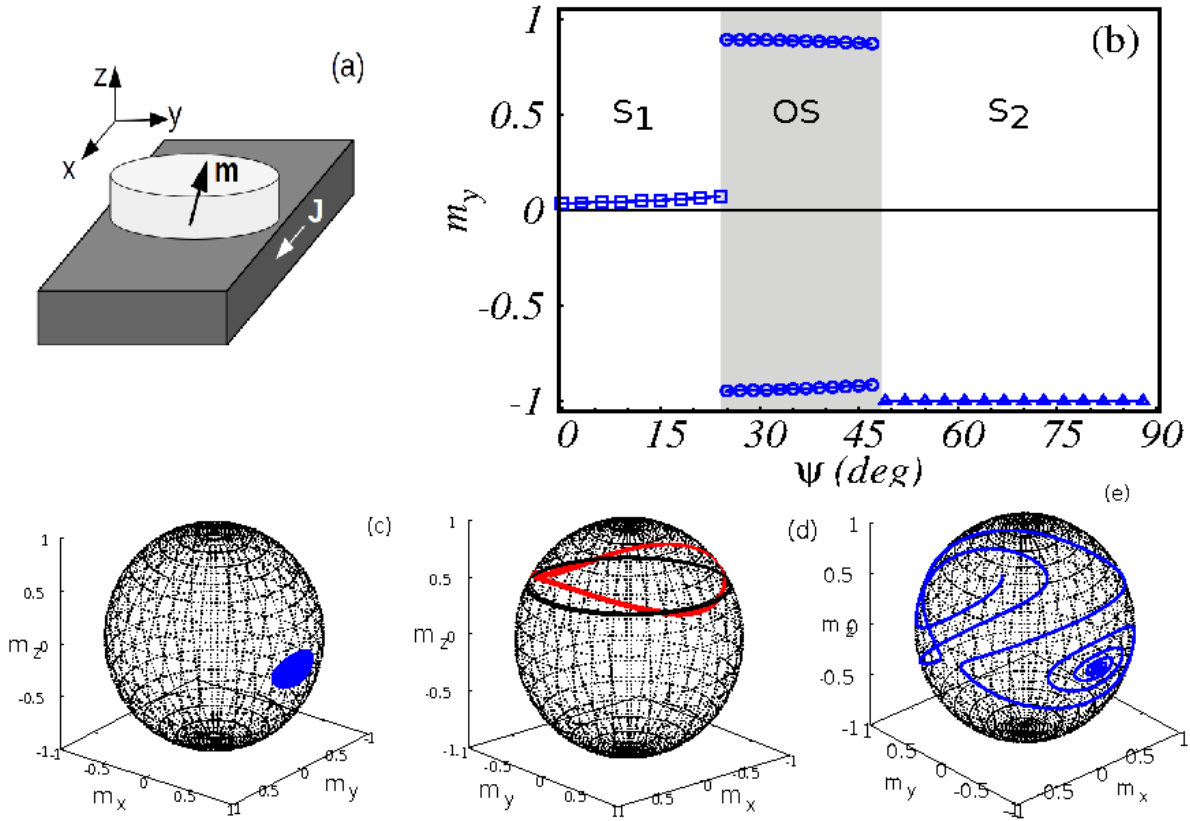}
	\caption{(Color online) (a) Schematic diagram of the spin-Hall system consisting a nonmagnetic layer(dark gray) and ferromagnetic layer(light gray).  (b) Bifurcation plot for maximum($m_y > 0$) and minimum($m_y < 0$) values of $m_y$ with respect to $\psi$. Magnetization trajectories corresponding to (c) steady state S$_1$ for $\psi$ = 20$^\circ$, (d) oscillations for $\psi$ = 25$^\circ$(red line) and 45$^\circ$(black line) and (e) steady state S$_2$ for $\psi$ = 60$^\circ$. Here, $\beta$ = 0.5, $J$ = 600 MA/cm$^2$ and $T$ = 0 K.}
	\label{fig1}
\end{figure}
To confirm the spin-Hall oscillations induced by field-like torque we consider a spin-Hall system as shown in Fig.\ref{fig1}(a), where a ferromagnetic layer is placed on a non-magnetic layer having high spin-orbit coupling.  The current flows along the $x$-direction in the nonmagnetic layer and the spins are polarized along the $y$-direction due to the spin-Hall effect and move along the $z$-direction into the ferromagnetic layer. The LLG equation that governs the magnetization dynamics of the ferromagnetic layer driven by spin-Hall torque is given by\cite{spicer:2018}

\begin{align}
\frac{d{\bf m}}{dt}=&-\gamma ~{\bf m}\times{\bf H}_{eff}+ \alpha ~{\bf m}\times\frac{d{\bf m}}{dt}\nonumber\\ &+\gamma H_{sh} ~{\bf m}\times ({\bf m}\times{\bf \hat{e}}_y)- \gamma \beta H_{sh}~{\bf m}\times{\bf \hat{e}}_y, \label{llg}
\end{align}

where ${\bf m} = m_x {\bf \hat{e}}_x+m_y {\bf \hat{e}}_y+m_z {\bf \hat{e}}_z$ is the unit vector of the magnetization, and ${\bf \hat{e}}_x,{\bf \hat{e}}_y, {\bf \hat{e}}_z$ are the unit vectors along $x,y,z$ directions respectively. $\gamma$ is the gyromagnetic ratio, $\alpha$ is the damping parameter, $\beta$ is the field-like torque parameter and $H_{sh}$ is the strength of the spin-Hall torque. The spin-Hall torque is given by
\begin{align}
H_{sh} = \frac {\hbar \theta_{sh} J}{ 2 e M_s d}, \nonumber
\end{align}
where $\hbar$ is the reduced Planck's constant, $\theta_{sh}$ is the spin-Hall angle, $J$ is the magnitude of current density, $e$ is the charge of electron, $M_s$ is the saturation magnetization and $d$ is the thickness of the ferromagnetic free layer.  The effective field ${\bf H}_{eff}$ is given by
\begin{align}
{\bf H}_{eff}=K_1~m_x~{\bf \hat{e}}_x+K_2~m_y~{\bf \hat{e}}_y-4\pi M_s m_z ~{\bf \hat{e}}_z+{\bf H}_{th}, \label{Heff}
\end{align}
where $K_1$ and $K_2$ are the two easy axis anisotropy fields along the $x$ and $y$ axis, respectively. $K_1$ and $K_2$ are varied from 0 to $K$ through the parameter $\psi$ by 
\begin{align}
K_1=K \cos\psi,~~~~~K_2=K \sin\psi, \nonumber
\end{align}
where
\begin{align}
\psi = \tan^{-1}(K_2/K_1). \nonumber
\end{align}

The term $4\pi M_s$ represents the demagnetization field and the thermal noise ${\bf H}_{th}$ is given by
\begin{align}
{\bf H}_{th} = \sqrt D~ {\bf G} \label{Hth},~~~D =   {\frac{2\alpha k_B T}{\gamma M_s \mu_0 V \triangle t}}.
\end{align}

Here, ${\bf G}$ is the Gaussian random number generator vector of the STNO with components $(G_{x}, G_{y}, G_{z})$, which satisfies the statistical properties $<G_{m}(t)>=0$ and $<G_{m}(t) G_{n}(t')>=\delta_{mn}\delta(t-t')$ for all $m,n=x,y,z$.  $k_B$ is the Boltzmann constant, $T$ is the temperature,  $V$ is the volume of the free layer, $\triangle t$ is the step size of the time scale used in the simulation and $\mu_0$ is the magnetic permeability at free space. The material parameters for the investigations are considered as $\gamma=1.764\times 10^7$ rad/(Oe s), $K=18.6$ kOe, $M_s=1448.3$ emu/c.c., $\alpha=0.005$, $\theta_{sh}=0.1$, $d=1$ nm and  $V$ = $\pi\times 50 \times 50 \times 1$ nm$^3$~\cite{tani:2017,tani:2018,tani:2015,kubota}.

\section{Results}

\subsection{Emergence of Steady and Oscillatory States}

For high power microwave voltage oscillations, large amplitude oscillations of the magnetization are expected.    Here, the oscillations are not exhibited for low current densities when the in-plane anisotropy is fixed along x or y axis.  However, when two easy axis anisotropies are formed along $x$ and $y$ directions, oscillations are found in the presence of field-like torque.  Since  out-of-plane oscillations are exhibited here, the oscillations of the magnetization over the parameter $\psi$ is verified by plotting the average of the extreme of $m_y$  against $\psi$.  To plot the out-of-plane oscillations of ${\bf m}$  Eq.\eqref{llg} is numerically solved by using the Runge-Kutta fourth-order method and the bifurcation plots for maximum and minimum values of $m_y$(blue) are plotted in Fig.\ref{fig1}(b) with respect to $\psi$ for $J$ = 600 MA/cm$^2$.  As indicated in Fig.\ref{fig1}(b), the system exhibits two steady state regions corresponding to the steady states S$_1$ and S$_2$(white regions), and one oscillatory region(OS). The steady states S$_1$ and S$_2$ are indicated by a line with squares and triangles respectively. The oscillatory state is indicated by a line with open circles.   In the steady state S$_1$, {\bf m} settles near the positive x-direction for both directions of current. In the steady state S$_2$, {\bf m} settles near the negative(positive) y-direction when the direction of current is positive(negative).  From Fig.\ref{fig1}(b) we can identify that the steady state regions S$_1$ and S$_2$ are exhibited below  $\psi = $25$^\circ$ and above $\psi = $48$^\circ$  respectively.  And the oscillatory region is exhibited between $\psi = $25$^\circ$ and $\psi = $48$^\circ$ and is found to be sandwiched between the two steady state regions. The trajectories of magnetization corresponding to the steady states S$_1$ and S$_2$ are plotted for $\psi$ = 20$^\circ$ and $\psi$ = 60$^\circ$ in Figs.\ref{fig1}(c) and \ref{fig1}(e) respectively. The trajectories corresponding to the oscillations are shown in Fig.\ref{fig1}(d) for 25$^\circ$ and 45$^\circ$ by red and black lines respectively, and these confirm the large amplitude out-of-plane oscillations of ${\bf m}$. 

From Fig.\ref{fig1}(b) we can see that the oscillations are not exhibited for the entire region of $\psi$.  Also, the oscillations are exhibited around 25$^\circ$ and disappear around 48$^\circ$. To understand this emergence and disappearance of oscillations with respect to $\psi$, the stability of the steady states S$_1$ and S$_2$ needs to be analyzed.

\subsection{Stability of Steady States $S_1$ and $S_2$}
To analyze the stability, the steady state points of S$_1$ and S$_2$ are identified as follows: Eq.\eqref{llg} is transformed into spherical polar coordinates using the equations $m_x = \sin\theta \cos\phi, ~m_y=\sin\theta\sin\phi$ and $m_z=\cos\theta$ as
\begin{align}
&\frac{(1+\alpha^2)}{\gamma}\frac{d\theta}{dt} = 2\pi M_s\alpha\sin 2\theta\nonumber\\ &+L_1(\alpha\cos\phi\cos\theta-\sin\phi)
+L_2(\alpha\sin\phi\cos\theta+\cos\phi)\nonumber\\
&+H_{sh}\left[(\alpha+\beta)\cos\phi-(1-\alpha\beta)\cos\theta\sin\phi\right],\label{spherical1}\\
&\frac{(1+\alpha^2)}{\gamma}\sin\theta\frac{d\phi}{dt} =-2\pi M_s\sin 2\theta\nonumber\\ &-L_1(\alpha\sin\phi+\cos\theta\cos\phi)
+L_2(\alpha\cos\phi-\cos\theta\sin\phi)\nonumber\\
&-H_{sh}\left[(1-\alpha\beta)\cos\phi+(\alpha+\beta)\cos\theta\sin\phi\right],\label{spherical2}\\
&\rm{where}\nonumber\\
&L_1 = K\cos\psi\sin\theta\cos\phi,~~L_2 = K\sin\psi\sin\theta\sin\phi\nonumber.
\end{align}
\begin{figure}[htp]
\centering\includegraphics[angle=0,width=1\linewidth]{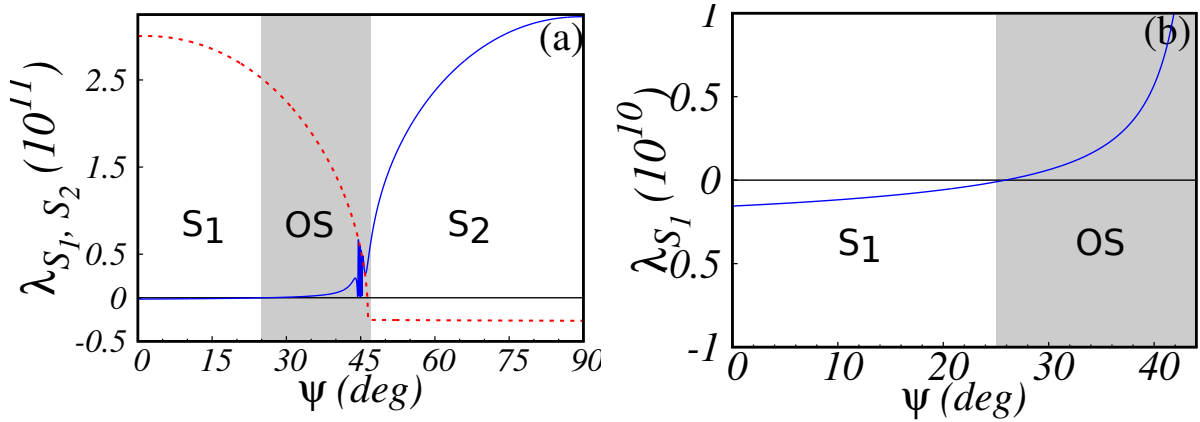}
\caption{(Color online) (a) Real parts of the maximum eigen values $\lambda_{S_1}$ and $\lambda_{S_2}$ corresponding to the steady states S$_1$(blue solid line) and S$_2$(red dotted line) respectively against $\psi$.  (b) Enlarged view for $\lambda_{S_1}$. Here $\beta$ = 0.5, $J$ = 600 MA/cm$^2$ and $T$ = 0 K.}
\label{fig2}
\end{figure}
From Figs.\ref{fig1}(b) and \ref{fig1}(c) one can speculate the steady state points $\theta^*$ and $\phi^*$ of S$_1$ as $\theta^*=\pi/2\pm\delta_1$ and $\phi^*=\pm\delta_2$ for $J  \gtrless 0$, where $\delta_1,\delta_2\approx 0$. By substituting $\theta=\theta^*$ and $\phi=\phi^*$ in Eqs.\eqref{spherical1} and \eqref{spherical2}, and using linear approximations we get
\begin{align}
\pm 4\pi M_s \alpha\delta_1&\pm K\cos\psi(\alpha\delta_1+\delta_2)\pm K\sin\psi\delta_2(\alpha\delta_2\delta_1-1)\nonumber\\&-H_{sh}[\alpha+\beta+(1-\alpha\beta)\delta_1\delta_2]=0 \label{a1},\\
\pm 4\pi M_s \alpha\delta_1&\pm K\cos\psi(\delta_1-\alpha\delta_2)\pm K\sin\psi\delta_2(\alpha+\delta_1\delta_2)\nonumber\\&-H_{sh}[1-\alpha\beta-(\alpha+\beta)\delta_1\delta_2]=0. \label{a2}
\end{align}
From Eqs.\eqref{a1} and \eqref{a2} we obtain
\begin{align}
\delta_1 =  \frac{\pm H_{sh}}{4\pi M_s+K\cos\psi} ,~ \delta_2 = \frac{\pm H_{sh}\beta}{K(\cos\psi-\sin\psi)\mp H_{sh}\delta_1}. \nonumber
\end{align}
Hence, the steady state points corresponding to S$_1$ are obtained as
\begin{align}
&\theta^* = \frac{\pi}{2} \pm \frac{H_{sh}}{4\pi M_s+K\cos\psi},\label{a5} \\ 
&\phi^* = \frac{\pm H_{sh}\beta}{K(\cos\psi-\sin\psi)-\frac{H_{sh}^2}{4\pi M_s+K\cos\psi}}.\label{a6}
\end{align}

Similarly, the steady state points corresponding to S$_2$ are identified as $\theta^* = \pi/2$ and $\phi^*=3\pi/2$ when $J>0$ and $\theta^* = \pi/2$ and $\phi^*=\pi/2$ when $J<0$ .  The eigenvalues corresponding to the steady states S$_1$ and S$_2$ are determined from the linear stability analysis. The maximum values of the real parts of the eigenvalues of the steady states S$_1$ and S$_2$  are  denoted as $\lambda_{S_1}$ and $\lambda_{S_2}$ respectively ( see Appendix A for details).  The values of $\lambda_{S_1}$ and $\lambda_{S_2}$  are plotted in Fig.\ref{fig2}(a) by the solid blue line and the red dotted line respectively, for $J$ = 600 MA/cm$^2$ and $\beta$ = 0.5. The background colours, namely white and gray, have been chosen in accordance with the Fig.\ref{fig1}(b). The values of $\lambda_{S_1}$ and $\lambda_{S_2}$ are negative and positive respectively below $\psi$ = 25$^\circ$ (see Fig.\ref{fig2}(b)). This implies that the steady state S$_1$ is stable below 25$^\circ$ while S$_2$ is unstable and consequently, the system settles into the steady state S$_1$ without exhibiting any oscillation below $\psi$ = 25$^\circ$ as shown in the left side white region of Fig.\ref{fig1}(b).  Similarly, the values of $\lambda_{S_1}$ and $\lambda_{S_2}$ are positive and negative respectively above $\psi$ = 48$^\circ$, which means that S$_1$ is unstable and S$_2$ is stable and the system settles into the steady state S$_2$ above $\psi$ = 48$^\circ$ as indicated in the right side white region of Fig.\ref{fig1}(b).  Also, from the gray region of Fig.\ref{fig2}(a) we can observe  that between $\psi$ = 25$^\circ$ and 48$^\circ$ both the values of $\lambda_{S_1}$ and $\lambda_{S_2}$ become positive which indicates that both the steady states become unstable and oscillations are exhibited between $\psi$ = 25$^\circ$ and 48$^\circ$ as confirmed in the gray region of Fig.\ref{fig1}(b).  Thus the emergence of the oscillations in the given region of $\psi$ is due to the fact that the real part of the eigenvalues of the both steady states become positive.

\subsection{Frequency and Power Spectral Density of the Oscillations}
It is essential that the frequency of the oscillations should be in the microwave range and tunable by the external parameters like current for the fruitful applications of microwave generation in the present spin Hall system for practical applications.  Though the oscillation frequency can be obtained from numerical results, explicit expression for this is also important for a better understanding of the underlying phenomenon. Here, the frequency is derived as follows:  From Eqs.\eqref{spherical1} and \eqref{spherical2} we can obtain
\begin{align}
& \frac{d\theta}{dt}=\frac{1}{(1-\alpha\beta)}\nonumber\\
&\left\{-(\alpha+\beta)\sin\theta\frac{d\phi}{dt} + \frac{K}{2}\gamma(\sin\psi-\cos\psi)\sin\theta\sin2\phi\right. \nonumber\\ 
&\left.-\gamma\beta\left[\frac{K}{2}\left(\cos\psi\cos^2\phi+\sin\psi\sin^2\phi\right)+2\pi M_s\right]\sin2\theta\right.\nonumber\\
&\left.-H_{sh}\gamma(1+\beta^2)\cos\theta\sin\phi\right\}. \label{spherical3}
\end{align} 
Since the variations of $m_z$ are small (see Fig.\ref{fig1}(d)) $\theta$ can be approximated as constant and ${d\phi}/{dt}=2\pi f={2\pi}/{T}$, where $f$ is the frequency and $T$ is the time period of the oscillation.  By approximating that $\theta=\theta_{J}$ is constant, the frequency can be derived by substituting the Eq.\eqref{spherical3} in the equation $\int_0^{nT}\frac{d\theta}{dt}~dt=0$\cite{tani:2018a,arun:2020,arun:ieee}, where $n$ is the number of oscillations, as
\begin{align}
f = \frac{\gamma\beta|\cos\theta_J|}{2\pi(\alpha+\beta)}\left[\frac{K}{2}(\cos\psi+\sin\psi)+4\pi M_s\right].\label{freq}
\end{align}
Here, Eq.\eqref{freq} clearly implies that the out-of-plane oscillations of the magnetization in the spin-Hall system are possible only when the field-like torque is present\cite{sode}. Also, it is confirmed that the oscillations are possible even in the absence of in-plane anisotropy ($K$ = 0) while the field-like torque is present (not shown here).

Therefore, it is understood that the field-like torque plays an important role for exciting the magnetization oscillations in an in-plane magnetized ferromagnetic layer in the presence of spin-Hall torque. Recently, few other types of effects such as  Rashba effect\cite{miron,Johan,Zhao},  Dzyaloshinskii-Moriya interaction\cite{yang:2015,yang:2017} and interlayer coupling\cite{Gonca} have been found as favourable in spin-Hall systems for the power enhancement, tunability of frequency, performance and synchronization of multiple oscillators. Our present study shows that field-like torque can be a potentially very important candidate to induce large amplitudes of oscillations with high frequency in spin-Hall systems as demonstrated below.

\begin{figure}[htp]
\centering\includegraphics[angle=0,width=1\linewidth]{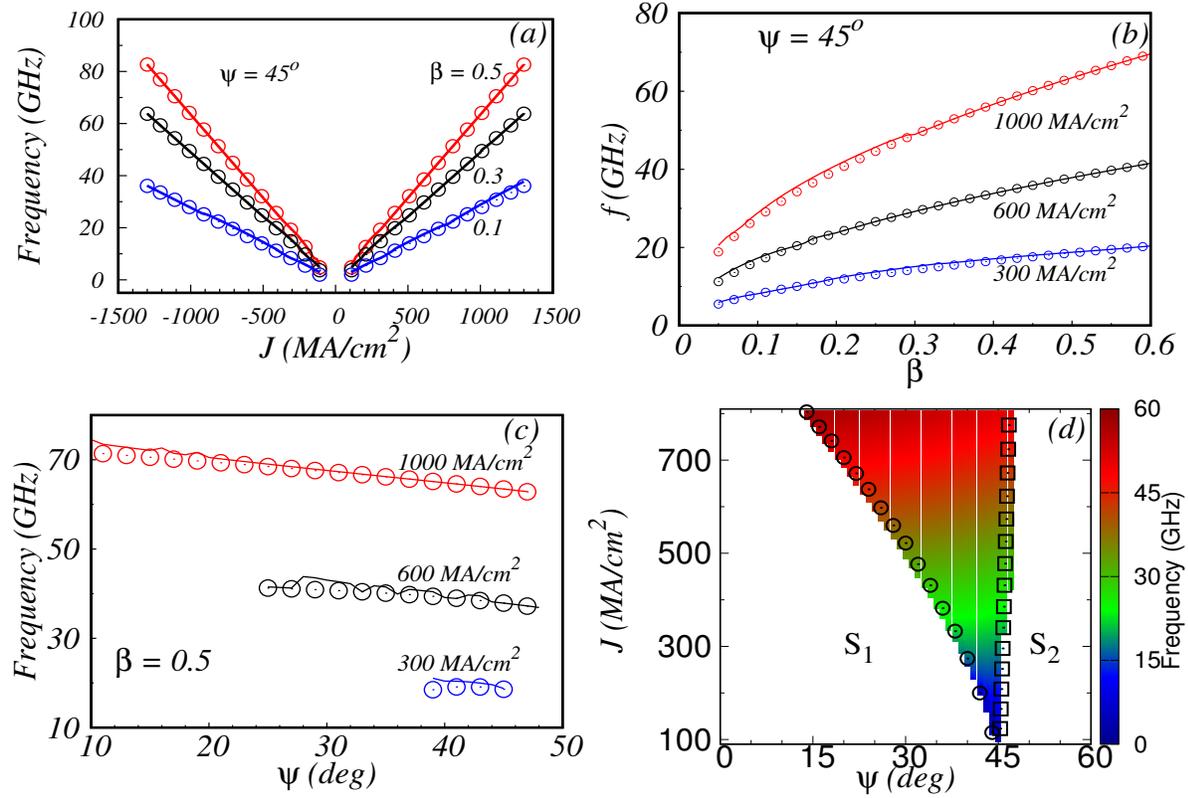}
\caption{(Color online) Oscillation frequency against (a) current density for $\beta$ = 0.1, 0.3 and 0.5 when $\psi$ = 45$^\circ$, (b) field-like torque for $J$ = 300, 600 and 1000 MA/cm$^2$  when  $\psi$ = 45$^\circ$, (c) $\psi$ for $J$ = 300, 600 and 1000 MA/cm$^2$  when  $\beta=0.5$. The open circles in figures (a), (b) and (c) are plotted for numerically computed frequencies and their corresponding lines are plotted by analytically computed frequencies from Eq.\eqref{freq}. (d) Frequency with respect to current density and $\psi$ for $\beta$ = 0.5. Here $T$ = 0 K.}
\label{fig3}
\end{figure}

The frequencies of oscillations of ${m_x}$ are plotted against current density, field-like torque and $\psi$ in Figs.\ref{fig3}(a), \ref{fig3}(b) and \ref{fig3}(c) respectively. In these figures, the numerically computed frequencies are indicated by open circles and their respective lines are plotted for the frequency obtained from the analytical expression, Eq.\eqref{freq}.  It can be seen that the analytically and numerically obtained frequencies match quite well, which confirms the reliability of numerical results.  Fig.\ref{fig3}(a) confirms the linear enhancement of the oscillation frequency with respect to current density when $\beta$ = 0.1, 0.3 and 0.5. The plots are symmetrical for negative and positive values of current density $J$, which implies the independent nature of the frequency on the direction of the current.   This is due to the fact that when $J$ is increased in the negative or positive direction the value of $\theta_J$ reaches 0$^\circ$ or 180$^\circ$ respectively. Therefore the value of $|\cos\theta_J|$ increases with the magnitude of $J$, and consequently the frequency increases with the magnitude of $J$.   Also, Fig.\ref{fig3}(a) confirms the large range of tunability in frequency from $\sim$2 GHz to $\sim$80 GHz. Further, the rate of change of frequency with respect to $J$ increases with the increase of field-like torque. The enhancement of the frequency by the field-like torque is confirmed in Fig.\ref{fig3}(b), where the frequency is plotted against field-like torque for different current densities $J$ = 300, 600 and 1000 MA/cm$^2$ for $\psi=45^\circ$. This confirms the existence of oscillations even for small values of the field-like torque. The existence of a range  of $\psi$ for oscillations with the increase of current density is confirmed in Fig.\ref{fig3}(c) by plotting the frequency against $\psi$ for $\beta$ = 0.5. Here we can also observe that the oscillation frequency can be maintained without any large drop when $\psi$ is varied. The two-parameter plot for frequency with respect to current density and $\psi$ is plotted for $\beta$ = 0.5 in Fig.\ref{fig3}(d) where we can identify the tunability and enhancement of frequency with respect to the current density. The steady state regions (white regions) S$_1$ and S$_2$ are indicated in Fig.\ref{fig3}(d).

\begin{figure}[htp]
	\centering\includegraphics[angle=0,width=1\linewidth]{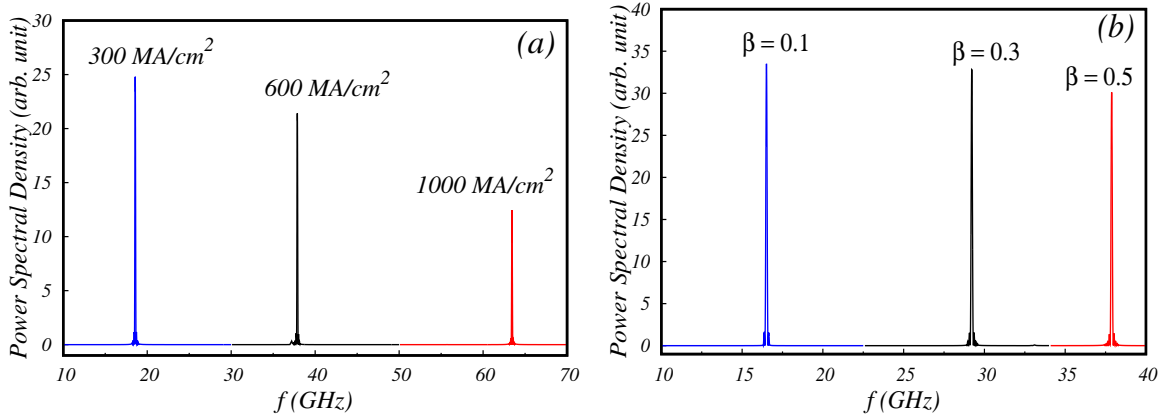}
	\caption{(Color online) Power spectral density of oscillations for different (a) current densities when $\beta$ = 0.5 and (b) field-like torques when $J$ = 600 MA/cm$^2$. Here $\psi$ = 45$^\circ$ and $T$ = 0 K.}
	\label{fig4}
\end{figure}

From Fig.\ref{fig3}(d) we can observe that there are two critical current densities $J_{S_1}$ and $J_{S_2}$ associated with the onset of oscillations near S$_1$ and offset of oscillations near S$_2$, respectively.  The value of $J_{S_1}$ is obtained from the linear stability condition for the fixed points \eqref{a5} and \eqref{a6} as (see Appendix A),
\begin{align}
&J_{S_1} = \pm \frac{d ~e}{\theta_{sh}^2\hbar^2} \sqrt{\frac{{ T_1-0.5 M_s^2 \sqrt{T_2 T_3 T_4 +T_5^2} }}{{\alpha+\beta}}}
\label{Jss1},
\end{align}
where $d$ is the thickness of the free layer and $e$ is the charge of the electron.~In Eq.(\ref{Jss1})
\begin{align}
T_1 = &(\alpha+\beta)K^2 M_s^2 + 32 (\alpha+2\beta-2\alpha\beta^2)M_s^4 \pi \nonumber\\
&+32(\alpha+\beta-\alpha\beta^2)K M_s^3 \pi \cos\psi + 2(2\alpha+\beta-2\alpha\beta^2)\nonumber\\&K^2 M_s^2 \cos^2 \psi 
+(\alpha+\beta)K^2 M_S^2 \cos 2\psi\nonumber\\&-16 K M_s^3 \pi\alpha\sin\psi-2K^2 M_s^2 \alpha \sin 2\psi,\nonumber\\
T_2 = &8\pi M_s\csc^2\psi-16\pi M_s\csc 2\psi\nonumber\\&+2K \csc\psi(-1+\cot 2\psi+\csc 2\psi),\nonumber\\
T_3 = &12 K M_s \pi \csc^2 \psi \csc 2\psi-8K \pi M_s \csc^2 2\psi \nonumber\\&+K^2 \csc^3 \psi + \csc\psi \csc 2\psi(-K^2+32\pi^2 M_s^2 \csc 2\psi),\nonumber\\
T_4 = &-8K\alpha(\alpha+\beta) \sin^2 \psi \sin^3 2\psi,\nonumber\\
T_5 = &2(\alpha+\beta)K^2 + 64 \pi^2 M_s^2 (\alpha+2\beta-2\alpha\beta^2)\nonumber\\&+2K^2(\alpha+\beta)\cos 2\psi-32KM_s\pi\alpha\sin\psi\nonumber\\&-4K^2\alpha\sin2\psi+32 K M_s\pi\alpha\csc\psi\sin2\psi(\alpha+\beta+\alpha\beta^2)\nonumber\\&+(2\alpha+\beta-2\alpha\beta^2)K^2\csc^2\psi\sin^22\psi.\nonumber
\end{align}
The calculated values of Eq.\eqref{Jss1} are indicated as open circles in Fig.\ref{fig3}(d).  The critical value of $\beta$ above which oscillations are possible can also be derived from the stability condition (A.4) (see Appendix A) as 
\begin{align}
\beta_c = - \frac{T_7}{\alpha} \left(-T_8 + \sqrt{T_8^2 - \left(\frac{\alpha}{H_{sh}}\right)^2 \frac{T_6+T_7}{2T_7}}\right),\label{betac}
\end{align}	
where
\begin{align}
&T_6 = 4\pi M_s +K \cos\psi, \nonumber\\
&T_7 = K (\cos\psi -  \sin\psi) - \frac{H_{sh}^2}{T_6}, \nonumber\\
&T_8 = \frac{1}{4T_6} + \frac{1}{2T_7} - \frac{K}{4 T_6 T_7}\left( \cos\psi-   \sin\psi\right). \nonumber
\end{align}

The value of $J_{S_2}$ is obtained from  linear stability condition (see Appendix A) for the fixed points $\theta^*  = \pi/2$ and $\phi^* = 3\pi/2$ and are plotted as open squares in Fig.\ref{fig3}(d). The form of $J_{S_2}$ is 
\begin{align}
&J_{S_2} = \frac{d~e~M_s \left( T_9-\sqrt{[(4\pi M_s \beta)^2+K(T_{10}+T_{11})]}\right)}{(1+\beta^2)\theta_{sh}\hbar},\label{Jss2}\\
\rm{where}\nonumber\\
&T_9 = \beta(4\pi M_s-K\cos\psi+2K \sin\psi),\nonumber\\
&T_{10} = K\beta^2 \cos^2\psi-4\sin\psi(4\pi M_s+K \sin\psi),\nonumber\\
&T_{11} = 4\cos\psi(2\pi M_s(2+\beta^2)+K \sin\psi).\nonumber
\end{align}
When the magnitude of the current density is below $J_{S_2}$, the steady state point $S_2$ becomes a stable focus\cite{Baza}. The coincidence of the critical current densities obtained from Eqs.\eqref{Jss1} and \eqref{Jss2} with the boundaries of oscillations near S$_1$ and S$_2$ shows the agreement between numerical and analytical results.

From Fig.\ref{fig2}(a) we can understand that oscillations are not possible when $\psi$ = 0$^\circ$ and 90$^\circ$ and they are possible essentially between 25$^\circ$ and 48$^\circ$.  These facts imply the need for two easy axis along $x$ and $y$ directions to make the spin-Hall oscillations. Also, from Eq.\eqref{betac} we can observe that oscillations can only be achieved above $\beta_c$. Thus, the steady state spin-Hall oscillations are executed due to the existence of two in-plane easy axis and the presence of field-like torque.

The impact of the current and field-like torque on the power of the oscillations is investigated by plotting the power spectral density for different values of $J$ and $\beta$ in Figs.\ref{fig4}(a) and \ref{fig4}(b) respectively in the absence of thermal noise. It is observed that both of the current density and field-like torque enhance the frequency substantially.  From Fig.\ref{fig4}(a) the Q-factor (which is the ratio between the frequency at the peak and its line-width) corresponding to $J$ = 300, 600 and 1000 MA/cm$^2$ are calculated as 211.07, 425.57 and 712.52 respectively. Similarly, from Fig.\ref{fig4}(b) the Q-factor corresponding to $\beta$ = 0.1, 0.3 and 0.5 are calculated as 187.59, 328.27 and 425.58 respectively. This implies that the Q-factor of the oscillations is increased by current as well as field-like torque.
\begin{figure}[htp]
	\centering\includegraphics[angle=0,width=1\linewidth]{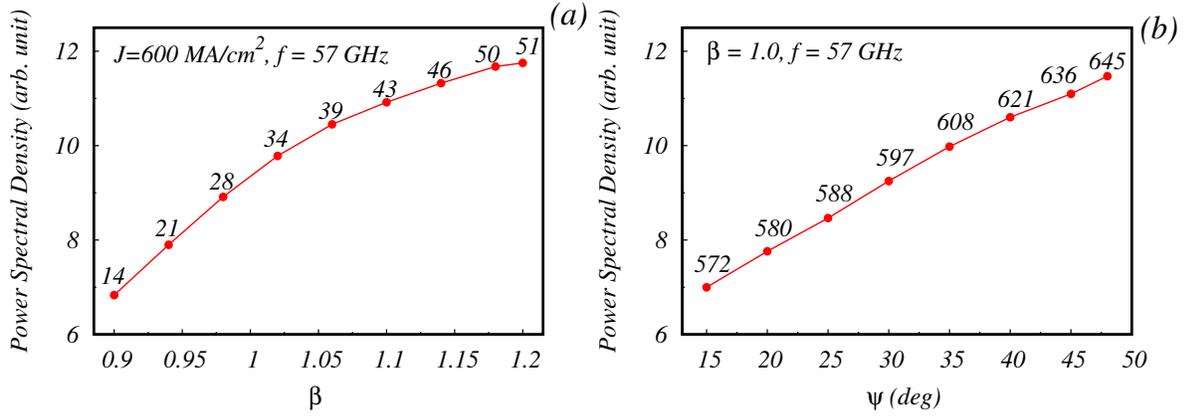}
	\caption{The power spectral density corresponding to the frequency 57 GHz (a) for different values of $\psi$ and $\beta$ when $J$ = 600 MA/cm$^2$ and (b) for different values of $J$ and $\psi$ when $\beta$ = 1.0.  Here $T$ = 0 K.}
	\label{pwr}
\end{figure}

Further, the variations in the power of the oscillations with constant frequency due to $\psi$ and current density are shown in Figs.\ref{pwr}(a) and (b), respectively.  The power spectral density corresponding to the frequency 57 GHz for different values of $\psi$ and $\beta$ is plotted in Fig.\ref{pwr}(a) when $J$ = 600 MA/cm$^2$, where the values of $\psi$ are given besides the respective data points inside the figure. It implies that the power can be enhanced by varying $\psi$ as we can see from the increment in the power spectral density to be nearly two times.  Similarly, Fig.\ref{pwr}(b) shows the power spectral density for the oscillations of frequency 57 GHz for different values of $J$ and $\psi$ when $\beta$ = 1.0. The values of $J$ are given besides the respective data points inside the figure.  It is clearly observed that the power spectral density can be enhanced again by current density to nearly two times. 
\subsection{Impact of Thermal Noise on Frequency and Power Spectral Density}

\begin{figure}[htp]
\centering\includegraphics[angle=0,width=1\linewidth]{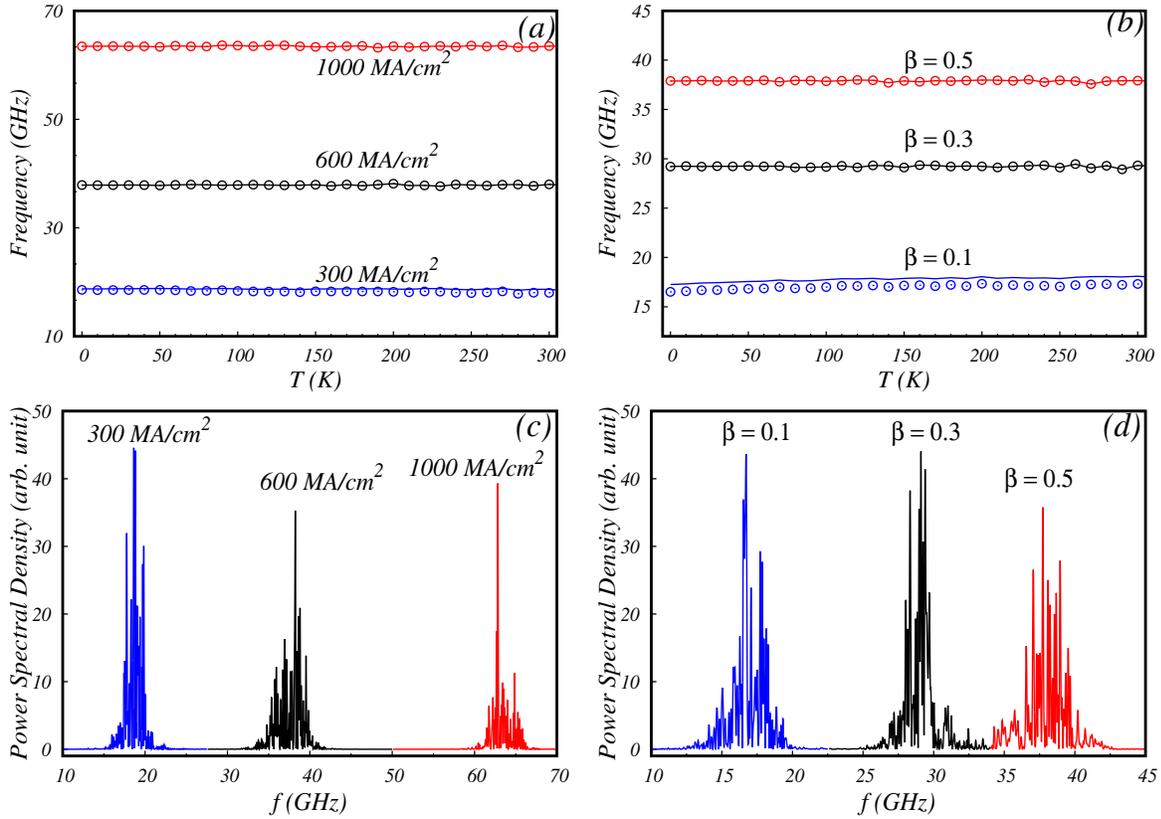}
\caption{(Color online) Frequency against temperature for  different  (a) current densities when $\beta$ = 0.5 and (b) field-like torques when $J$ = 600 MA/cm$^2$.  Power spectral density of oscillations for different (c) current densities when $\beta$ = 0.5 and (d) field-like torques when $J$ = 600 MA/cm$^2$ for $T$ = 300 K. Here $\psi$ = 45$^\circ$.}
\label{fig5}
\end{figure}

For realistic applications, the system should be stable against thermal noise when the current is increased. Here to investigate the stability of the system over thermal noise, the frequency of the oscillations is plotted against temperature in Figs.\ref{fig5}(a) and (b) for different values of $J$ (300, 600 and 1000 MA/cm$^2$) and $\beta$ (0.1, 0.3 and 0.5) respectively. The lines in Fig.\ref{fig5}(a) correspond to numerically computed frequencies and the open circles correspond to the frequencies computed from Eq.\eqref{freq}.  Figs.\ref{fig5}(a) and (b) confirm that an increase of temperature does not affect the oscillation frequency by an appreciable amount as the magnitudes of current and field-like torque increase. Also, the power spectral density is plotted for different values of current density and field-like torque in Figs.\ref{fig5}(c) and (d) respectively in the presence of temperature $T$ = 300 K. By comparing Figs.\ref{fig5}(c) and (d) with Figs.\ref{fig4}(a) and (b) we can understand that apart from the additional spikes due to noise there is no shift in the frequencies due to the presence of temperature.

\section{Conclusions}
In summary, the emergence of magnetization oscillations in spin-Hall oscillator by tuning the direction of in-plane anisotropy direction in the presence of field-like torque is confirmed without any external field and polarizer by analytical and numerical investigations on the LLG equation. It is observed that the range of $\psi$ for which the oscillations are possible increases with the increase of current in the presence of field-like torque. We have also shown the existence of large amplitude out-of-plane oscillations and derived the oscillation frequency analytically which matches well with the numerically computed frequencies. Further, we have analytically proved that the field-like torque is essential for spin-Hall oscillations.  It is shown that the frequency can be enhanced substantially by tuning the current, field-like torque and $\psi$.  The frequency can be tuned from $\sim$2 GHz to $\sim$80 GHz by the current.    Also, the enhancement of frequency by current and field-like torque along with the enhancement of the Q-factor is confirmed. Further, the stability of the system against thermal noise is proved.

\section*{Appendix A}
\subsection*{Stability condition and critical current densities}
The stability of the fixed points $S_1$ and $S_2$ is derived from the eigenvalues of the Jacobian matrix of the system associated with equations \eqref{spherical1} and \eqref{spherical2} at the steady state $(\theta^*,\phi^*)$, which is given by
$$
M = \begin{pmatrix}
\frac{dP}{d\theta}|_{(\theta^*,\phi^*)}  & \frac{dP}{d\phi}|_{(\theta^*,\phi^*)}\\
\frac{dQ}{d\theta}|_{(\theta^*,\phi^*)}  &  \frac{dQ}{d\phi}|_{(\theta^*,\phi^*)} 
\end{pmatrix}.
\qquad \qquad \qquad  ({\rm A}.1)
$$
Here $P(\theta,\phi)$ and $Q(\theta,\phi)$ are defined from  equations \eqref{spherical1} and \eqref{spherical2} as
\begin{align}
&P(\theta,\phi) = \frac{\gamma}{(1+\alpha^2)}\left\{2\pi M_s\alpha\sin 2\theta\right.\nonumber\\
&\left.+L_1(\alpha\cos\phi\cos\theta-\sin\phi)+L_2(\alpha\sin\phi\cos\theta+\cos\phi)\right.\nonumber\\&\left.+H_{sh}\left[(\alpha+\beta)\cos\phi-(1-\alpha\beta)\cos\theta\sin\phi\right]\right\},\tag{A.2a}\label{Appx1}\\
&Q(\theta,\phi) =\frac{\gamma \csc\theta}{(1+\alpha^2)}\left\{-2\pi M_s\sin 2\theta\right.\nonumber\\
&\left.-L_1(\alpha\sin\phi+\cos\theta\cos\phi)+L_2(\alpha\cos\phi-\cos\theta\sin\phi)\right.\nonumber\\&\left.-H_{sh}\left[(1-\alpha\beta)\cos\phi+(\alpha+\beta)\cos\theta\sin\phi\right]\right\}.\tag{A.2b}\label{Appx2}
\end{align}
The stability nature of the steady state points is determined from the following eigenvalues of the matrix $M$\cite{Lakshmanan}, 
\begin{align}
\lambda_{\pm}=\frac{1}{2}\left[{\rm Tr}(M)\pm\sqrt{{\rm Tr}(M)^2-4~{\rm Det}(M)}\right] .\tag{A.3}\label{app1}
\end{align}
To   understand the stability of the steady states, the largest real part of the eigenvalues $\lambda_{+}$ and $\lambda_-$ (here Re$[\lambda_{+}] >$ Re$[\lambda_-]$) is calculated for both the steady states S$_1$ ($\lambda_{S_1}$) and S$_2$ ($\lambda_{S_2}$), and plotted in Figs.\ref{fig2}.  We have identified that while varying  $\psi$ the stability of the fixed point corresponding to $S_1$ ($S_2$) state changes from stable spiral (saddle) to unstable spiral (stable spiral).  The conditions from which the stability of the steady states S$_1$ and S$_2$ can be analyzed are derived from the eigenvalue equations Eq.\eqref{app1} and are given by 
\begin{align}
{\rm Tr}(M)&=0\quad \mbox{for} \quad {\rm S}_1,& \tag{A.4} \label{app3}\\
{\rm Det}(M)&=0 \quad \mbox{for} \quad {\rm S}_2.& \tag{A.5} \label{app4}
\end{align}
The critical current densities $J_{S_1}$ and $J_{S_2}$ given in Eqs.\eqref{Jss1} and \eqref{Jss2} are derived from the stability conditions \eqref{app3} and \eqref{app4} respectively.
\section*{Appendix B}
\subsection*{Frequency of precession}
To find the frequency of the oscillations the following equation is obtained by adding the Eqs.\eqref{spherical1} and \eqref{spherical2} after multiplying them with $(1-\alpha\beta)$ and $(\alpha+\beta)$ respectively.
\begin{align}
(1-\alpha\beta)\frac{d\theta}{dt} + (\alpha+\beta) \sin\theta \frac{d\phi}{dt} =&-2\pi M_s \gamma \beta \sin2\theta - K \gamma \cos\psi \sin\theta \cos\phi (\beta \cos\theta \cos\phi+\sin\phi) 
\nonumber\\
&+ K\gamma \sin\psi \sin\theta\sin\phi(\cos\phi-\beta \sin\phi\cos\theta)\nonumber\\
& - H_{sh} \gamma (1+\beta^2)\cos\theta\sin\phi. \tag{B.1}\label{B1}
\end{align}
From Eq.\eqref{B1} we can derive,
\begin{align}
\frac{d\theta}{dt} = \frac{1}{1-\alpha\beta} \left\{-2\pi M_s \gamma\beta \sin2\theta - K\gamma \cos\psi\sin\theta\cos\phi(\beta\cos\theta\cos\phi+\sin\phi)\right.\nonumber\\\left.+K\gamma\sin\psi\sin\theta\sin\phi(\cos\phi-\beta\sin\phi\cos\theta)\right.\nonumber\\\left.-H_{sh}\gamma(1+\beta^2)\cos\theta\sin\phi-(\alpha+\beta)\sin\theta \frac{d\phi}{dt} \right\}\nonumber,
\end{align}

\begin{align}
=  \frac{1}{1-\alpha\beta}\left\{-\gamma \beta\left[2\pi M_s+\frac{K}{2}(\cos\psi\cos^2\phi+\sin\psi\sin^2\phi) \right]\sin2\theta\right.\nonumber\\\left.+\frac{K}{2} \gamma (\sin\psi-\cos\psi)\sin\theta\sin2\phi \right.\nonumber\\\left.-H_{sh} \gamma (1+\beta^2)\cos\theta\sin\phi -(\alpha+\beta)\sin\theta\frac{d\phi}{dt}  \right\}.\tag{B.2} \label{B2}
\end{align}
Once the magnetization reaches steady state oscillation, it is assumed that the polar angle of the magnetization precession $\theta$ is constant as $\theta_{J}$ and therefore $d\theta/dt=0$.  Since $\theta=\theta_J$, the rate of change of azimuthal angle $d\phi/dt = \omega$, where $\omega$ is the angular velocity of the precession.    By integrating Eq.\eqref{B2}, with respect to time from 0 to $nT$ for $n$ number of oscillations, we can get
\begin{align}
\int_0^{nT}\frac{d\theta_J}{dt}dt = \frac{1}{1-\alpha\beta}\int_0^{nT} \left\{-\gamma \beta\left[2\pi M_s+\frac{K}{2}(\cos\psi\cos^2\phi+\sin\psi\sin^2\phi) \right]\sin2\theta_J\right.\nonumber\\\left.+\frac{K}{2} \gamma (\sin\psi-\cos\psi)\sin\theta_J\sin2\phi\right.\nonumber\\\left. -H_{sh} \gamma (1+\beta^2)\cos\theta_J\sin\phi -(\alpha+\beta)\sin\theta_J \frac{d\phi}{dt}  \right\}dt.\tag{B.3} \label{B3}
\end{align}
Since $\frac{d\theta_j}{dt}=0$, Eq.\eqref{B3} can be written as
\begin{align}
&\int_0^{nT} \left\{-\gamma \beta\left[2\pi M_s+\frac{K}{2}(\cos\psi\cos^2\phi+\sin\psi\sin^2\phi) \right]\sin2\theta_J\right.\nonumber\\&\left.+\frac{K}{2} \gamma (\sin\psi-\cos\psi)\sin\theta_J\sin2\phi -H_{sh} \gamma (1+\beta^2)\cos\theta_J\sin\phi -(\alpha+\beta)\sin\theta_J \frac{d\phi}{dt}  \right\}dt =0.\tag{B.4}\label{B4}
\end{align}

Since $d\phi/dt=\omega$, Eq.\eqref{B4} is integrated with respect to $\phi$ by substituting $dt = d\phi/\omega$ and we get
\begin{align}
&\int_0^{2n\pi} \frac{1}{\omega}\left\{-\gamma \beta\left[2\pi M_s+\frac{K}{2}(\cos\psi\cos^2\phi+\sin\psi\sin^2\phi) \right]\sin2\theta_J\right.\nonumber\\&\left.+\frac{K}{2} \gamma (\sin\psi-\cos\psi)\sin\theta_J\sin2\phi -H_{sh} \gamma (1+\beta^2)\cos\theta_J\sin\phi -(\alpha+\beta)\omega \sin\theta_J \right\}d\phi =0.\tag{B.5} \label{B5}
\end{align}
After integrating Eq.\eqref{B5} we can get
\begin{align}
\gamma \beta \cos\theta_J \left[4\pi M_s + \frac{K}{2}(\cos\psi+\sin\psi)\right]+(\alpha+\beta)\omega = 0 ,\tag{B.6}\label{B6}
\end{align}
From Eq.\eqref{B6} we can derive the angular velocity
\begin{align}
\omega = -\frac{\gamma\beta\cos\theta_J}{\alpha+\beta}\left[4\pi M_s + \frac{K}{2}(\cos\psi+\sin\psi)\right],\tag{B.7}\label{B7}
\end{align}
and the frequency is
\begin{align}
f = \frac{\gamma\beta|\cos\theta_J|}{2\pi(\alpha+\beta)}\left[4\pi M_s + \frac{K}{2}(\cos\psi+\sin\psi)\right].\tag{B.8}\label{B8}
\end{align}

\section*{Acknowledgements}
The work of V.K.C. forms part of a research project sponsored by DST-SERB-CRG Project No. CRG/2020/004353. M.L. wishes to thank the Department of Science and Technology for the award of a SERB Distinguished Fellowship under Grant No.SB/DF/04/2017 in which R. Arun is supported by a Research Associateship.\\~\\

\end{document}